\documentstyle[11pt,newpasp,twoside,epsf,epsfig]{article}
\def\simless{\mathbin{\lower 3pt\hbox
     {$\rlap{\raise 5pt\hbox{$\char'074$}}\mathchar"7218$}}}   %< or of order
\def\simmore{\mathbin{\lower 3pt\hbox
     {$\rlap{\raise 5pt\hbox{$\char'076$}}\mathchar"7218$}}}   %> or of order
\def\edcomment#1{\iffalse\marginpar{\raggedright\sl#1\/}\else\relax\fi}
\reversemarginpar

\begin{document}
\title{First observation of a transition between ``parallel
tracks'' in the kHz QPO frequency vs. intensity diagram}
 \author{Mariano M\'{e}ndez}
\affil{SRON, National Institute for Space Research, Sorbonnelaan 2, 3584
CA Utrecht, The Netherlands}

\begin{abstract}
Contrary to theoretical expectations, observations with the {\em Rossi
X-ray Timing Explorer (RXTE)} show that in X-ray binaries timing
properties are not uniquely correlated with X-ray luminosity. For
instance, although the frequencies of the kilohertz quasi-periodic
oscillations (kHz QPOs) correlate with X-ray flux on short ($\sim$few
hours) time scales, on time scales longer than a day the QPO appears at
more or less the same frequency, whereas the luminosity may be a factor
of a few different. The result is a set of almost parallel tracks in a QPO
frequency vs. X-ray flux plot. Despite the ``parallel tracks'' are a
common phenomenon among kHz QPO sources, until now, after five years of
observations with RXTE, not a single transition between two of these
tracks had been seen. Here I present the first detection of such a
transition, in 4U\,1636--53.
\end{abstract}

\section{Introduction}

Observations with the {\em Rossi X-ray Timing Explorer (RXTE)} have
revealed kilohertz quasi-periodic oscillations (kHz QPOs) in some 20
X-ray binaries. It is generally thought that these kHz QPOs reflect the
motion of matter in orbit at some preferred radius in the accretion
disk around the neutron star (see van der Klis 2000 for a review of the
phenomenology of these QPOs, and for a description of the models so far
proposed to explain them).

Calculations by Miller, Lamb, \& Psaltis (1998) show that the inner
radius of the disk is set by angular momentum losses to the radiation
field: When mass flow through the disk increases, the inner radius of
the disk decreases, and therefore the QPO frequency (the Keplerian
frequency at this radius) increases. But because in accretion-powered
systems luminosity is proportional to the mass accreted onto the
compact object, from the above it follows that there should be a
one-to-one relation between QPO frequency and bolometric flux. 

To the extent that X-ray flux is a good measure of the bolometric flux
(see, e.g., Ford et al. 2000), observations with RXTE seem to
contradict the above expectations. Figure 1a shows a plot of QPO
frequency vs. X-ray intensity for the transient source 4U\,1608--52
(M\'endez et al. 1999). Each segment there represents an uninterrupted
observation lasting $\simless$1 hour, whereas different tracks denote
observations separated by intervals longer than a day.  From this
Figure it is apparent that frequency and X-ray count rates are
positively correlated during relatively short periods, but they are
uncorrelated over longer time intervals.

\section{Results and Discussion}

Despite 5 years of observations with RXTE, so far no single transition
between two tracks in a QPO frequency vs. X-ray intensity diagram had
been observed. Here I report the first direct observation of one such
transition, in an RXTE observation of 4U\,1636--53 made in 1998 (see
Figure 1b). The observation starts with the source at the upper end of
``Track 1'' (``BEGIN''), and ends at the upper end of ``Track 2''
(``END''). In between, 4U\,1636--53 goes three times from one track to
the other, as indicated by the arrows with numbers 1, 2, and 3. The
transitions are very fast ($\simless$300 s), consistent with upper
limits inferred from the time intervals between two consecutive tracks
in this and other sources, in diagrams similar to the one shown in
Figure 1a (e.g., M\'endez 2000).

\begin{figure}
%\centerline{
\plotfiddle{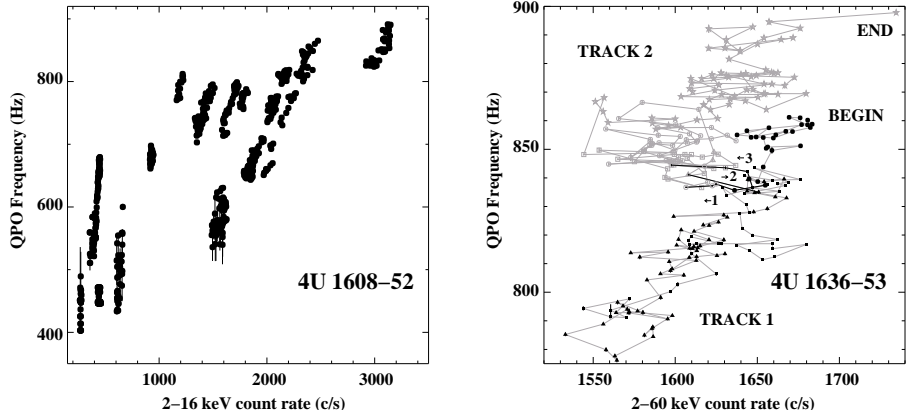}{5.15cm}{0}{30}{30}{-160}{0}
%}
%\centerline{\epsfig{file=mendez_fig1.eps,height=5.1cm}}
\caption{(a) QPO frequency vs. X-ray intensity for 4U\,1608--52, showing
the so-called ``parallel tracks''. (b) Transition between two tracks 
(gray and black points) in 4U\,1636--53.}
\end{figure}

During the transitions in 4U\,1636--53, X-ray intensity changes by
$\simless$3\,\%. From Figure 1a we see that in 4U\,1608--52 X-ray
intensity differs by factors of a few between tracks (in 4U\,1636--53
jumps of $\sim$30\,\% are observed over longer time intervals than
shown in Figure 1b), it is therefore not clear whether this larger
differences in X-ray intensity are the cumulative effect of several
small jumps, or if they have a completely different origin. 

I will present these results in more detail elsewhere (M\'endez \& van
der Klis, 2002, in preparation), and there I will discuss more
extensively their implications upon models of the accretion flow in
X-ray binaries.

%\acknowledgments


\begin{references}
\reference Ford, E., et al. 2000, \apj, 537, 368
\reference M\'endez, M., et al. 1999, \apj, 511, L49
\reference M\'endez, M. 2000, 19th. Texas Symp, 15/16
%\reference M\'endez, M., \& van der Klis, M. 2001, in prep.
\reference Miller, C., Lamb, F. K., \& Psaltis, D. 1998, \apj, 508, 791
\reference van der Klis, M. 2000, \araa, 38, 717
\end{references}
\end{document}